\documentclass[twocolumn,
trackchanges,twocolappendix]{aastex63}
\usepackage{xcolor}

\DeclareRobustCommand{\ers}{\bgroup\markoverwith{\textcolor{red}{\rule[.5ex]{2pt}{0.4pt}}}\ULon}

\received{}
\revised{}
\accepted{}
\submitjournal{ApJ}
\shorttitle{Shallow Decay Phase of GRB Afterglows}
\shortauthors{Asano}
\begin{document}

\title{
Multiwavelength Modeling for the Shallow Decay Phase of Gamma-Ray Burst Afterglows
}

\author{Katsuaki Asano}
\affiliation{Institute for Cosmic Ray Research, The University of Tokyo, 5-1-5 Kashiwanoha, Kashiwa, Chiba 277-8582, Japan}

\begin{abstract}
We simulate the emission in the shallow decay phase of gamma-ray burst afterglows using a time-dependent code. We test four models: the energy injection model, evolving the injection efficiency of non-thermal electrons, evolving the amplification of the magnetic field, and the wind model with a relatively low bulk Lorentz factor. All of the four models can reproduce the typical X-ray afterglow lightcurve.
The spectral shape depends on not only the parameter values at the time corresponding to the observer time but also the past evolution of the parameters. The model differences appear in the evolution of the broadband spectrum, especially in the inverse Compton component. Future gamma-ray observations with imaging atmospheric Cherenkov telescopes such as CTA will reveal the mechanism of the shallow decay phase.
\end{abstract}

\section{Introduction\label{sec:intro}}

A large fraction of gamma-ray bursts (GRBs) show X-ray afterglows with a decay index of $<0.5$ during initial $\sim 10^3$ s \citep{2006ApJ...642..389N,2007ApJ...670..565L}. This shallow decay phase (or plateau phase) is one of the unsolved problems in the GRB afterglow physics. One of the most frequently discussed models is the energy injection model \citep{2001ApJ...552L..35Z,2006ApJ...642..354Z,2006MNRAS.366L..13G}, in which the energy in the forward shock region increases with time, so that the decay of the bulk Lorentz factor becomes slower than the standard Blandford--McKee solution \citep{1976PhFl...19.1130B}. However, other type of models can be possible; for example, the two-component jets or anisotropic jet structure \citep{2006ApJ...641L...5E,2006ApJ...640L.139T,2006MNRAS.370.1946G}, off-axis jet \citep{2020MNRAS.492.2847B}, evolving microphysical parameters \citep{2006MNRAS.369..197F,2006MNRAS.369.2059P,2006A&A...458....7I}, thick shell model \citep{2014MNRAS.437.2448L}, and the decreasing circumstellar density \citep[wind model;][]{2012ApJ...744...36S,2022NatCo..13.5611D}.

Optical afterglows frequently show a single power-law decay \citep{2008MNRAS.387..497P,2015ApJ...805...13L}, while the simultaneous X-ray afterglow is shallower.
The early optical afterglow may be dominated by the reverse shock emission; thus optical observations may be not an ideal method to investigate the mechanism of the shallow decay phase.

GeV--TeV gamma-ray observations can provide clues to the shallow decay phase. GeV afterglows in initial $10^3$--$10^4$ s are detected with {\it Fermi}-LAT for a fraction of GRBs \citep{2019ApJ...878...52A}.
However, \citet{2020MNRAS.494.5259Y} claimed that GRBs detected with {\it Fermi}-LAT tend not to show the X-ray shallow decay behavior. This may suggest that much deeper observations in the GeV band are needed to detect GeV afterglows accompanying a shallow-decaying X-ray afterglow.

Recently, imaging atmospheric Cherenkov telescopes have detected TeV afterglows from several GRBs \citep{2019Natur.575..459M,2019Natur.575..464A,2021Sci...372.1081H,2024MNRAS.527.5856A}. The detection of more than 64,000 photons above 0.2 TeV from GRB 221009A with LHAASO \citep{2023Sci...380.1390L} is encouraging for future observations of TeV afterglows.

In this paper, we simulate broadband emission from the shallow decay phase with several models using our time-dependent code. The model differences appear in the evolution of the photon spectra, especially in the inverse Compton (IC) component. Future gamma-ray observations of GRBs will reveal the mechanism of the shallow decay phase.

This paper is organized as follows. 
In \S \ref{sec:method}, we explain our method to simulate the afterglow emission. We test four models for the shallow decay phase, which are presented in \S \ref{sec:model}.
The results for the four models are shown in \S \ref{sec:results}.
\S \ref{sec:sum} is devoted to summary.

\section{Method}\label{sec:method}

In this paper, to simulate the evolution of GRB afterglow spectra, we adopt the numerical code in \citet{2017ApJ...844...92F}, which was also used in \citet{2020ApJ...905..105A}. The code is one-zone time-dependent to simulate the evolution of the electron and photon energy distributions in the emission region, which is the shocked circumstellar medium (CSM). The physical processes included in this code are particle injection, synchrotron emission, synchrotron self-absorption, IC emission, $\gamma \gamma$-absorption, electron--positron pair creation, adiabatic cooling, and photon escape. With the energy conservation law, the bulk Lorentz factor $\Gamma$ of the emission region is developed with the radius $R$ from the central engine taking into account mass loading and photon emission. The emission region is treated as a uniform shell. The shock jump condition provides the density $\rho_{\rm sh}$ just behind the shock front, with which the one-zone approximation provides the width $W$ of the emission region as $4 \pi \int \rho_{\rm CSM} r^2 dr=4 \pi R^2 W \rho_{\rm sh}$, where $\rho_{\rm CSM}$ is the CSM density.

Electrons are injected with a power-law index $p$ ($\dot{n}_{\rm e} \propto \varepsilon_{\rm e}^{-p}$) with the energy fraction $\epsilon_{\rm e}$ to the injected energy into the shell. Those two parameters regulate the minimum Lorentz factor $\gamma_{\rm min}$ of electrons at injection. The magnetic field evolves with the energy fraction $\epsilon_B$. See \citet{2017ApJ...844...92F} for other details. The photon spectrum for an observer is calculated consistently with the amount of escaped photons. Assuming a spherically symmetric geometry, the code integrates photons escaped from the emission surface, taking into account the Doppler beaming and the surface curvature to address the photon arrival time.

Most of the analytical formulae for GRB afterglow spectra \citep{1998ApJ...497L..17S,2000ApJ...543...66P,2001ApJ...552L..35Z} are based on the approximation that the electron energy distribution is instantly determined by the local quantities such as $\Gamma$ and the CSM density at that radius $R$, irrespectively of the past evolution of the emission region. Our time-dependent code is free from this instant control approximation and one-to-one correspondence approximation between $R$ and the observer time $t_{\rm obs}$ \citep[see, e.g.][]{2021ApJ...923..135D}.

In this paper, we change the photon escape process in the numerical code, which affects the photon energy density. Isotropic photons in the shell escape from both the front and rear surfaces. In \citet{2017ApJ...844...92F}, the boundary of the shell is assumed to be static in the shell rest frame, so that the photon escape rate is $c N_\gamma/(2 W)$, where $N_\gamma$ is the total photon number in the shell. However, as the shock is relativistic in early afterglows, the shock front in the shocked fluid rest frame propagates with a speed of $c/3$. Only photons with the angle $\theta$ of $\cos\theta>1/3$ can escape with the effective velocity of $v_{\rm e}=c (\cos\theta-1/3)$. As the photon number density per unit $\mu=\cos\theta$ in the isotropic case is $n_\gamma/2$, where $n_\gamma=N_\gamma/(4 \pi R^2 W)$ in the one-zone approximation, the photon escape rate from the forward shock front is $4 \pi R^2 \int_{1/3}^{1} d\mu v_{\rm e} n_\gamma/2$. The rear boundary is the contact discontinuity or rarefaction wave in actual cases \citep[e.g.][]{2023MNRAS.526..512K}, so the boundary velocity may be non-relativistic. However, such an asymmetric structure makes it difficult to estimate the photon density consistently in the one-zone and isotropic approximation. For simplicity, we adopt the same escape rate even for the rear boundary, and then the escape rate becomes 
\begin{eqnarray}
\dot{N}_\gamma=\frac{2cN_\gamma}{9W}.
\end{eqnarray}

\begin{figure}
    \centering
    \epsscale{1.2}
    \plotone{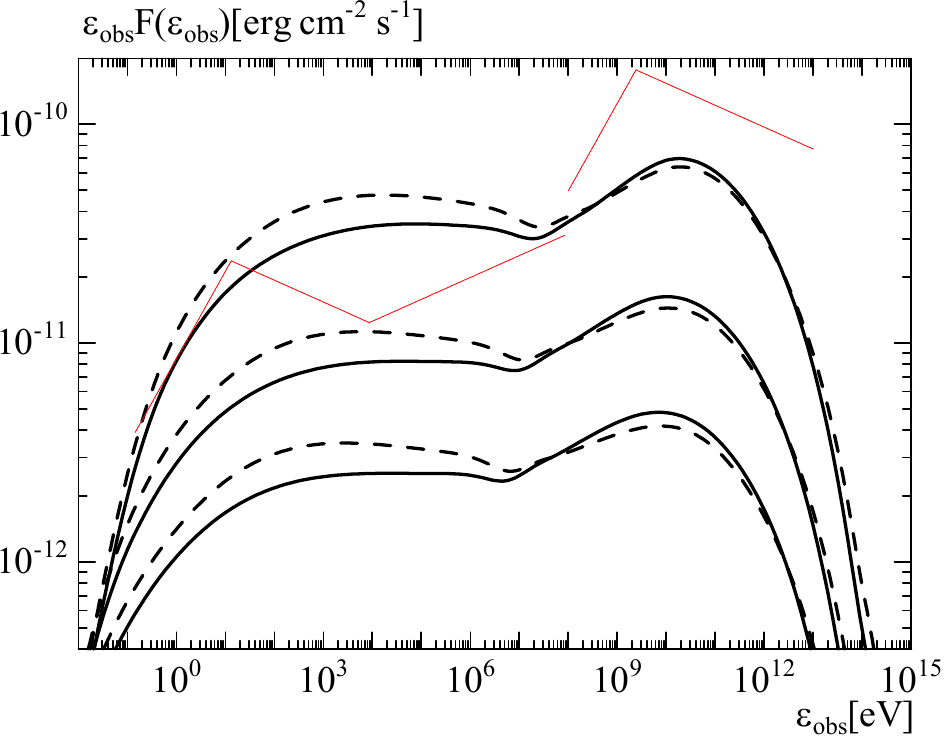}
    \caption{Comparison of deabsorbed photon spectra (neglecting EBL absorption) between the previous (dashed) and present (solid) methods for the photon escape process. From top to bottom, the observed time is 3000 s, 10,000 s, and 30,000 s, respectively. The parameters are $z=0.5$, $E_0=10^{53}$ erg, $\Gamma_0=300$, $n_0=1~\mbox{cm}^{-3}$, $\epsilon_{\rm e}=0.1$, $\epsilon_B=10^{-3}$, and $p=2.2$. The red solid lines are the analytical spectrum based on the formulae in \citet{2009ApJ...703..675N} for the time 3000 s.}
    \label{fig:Method}
\end{figure}

We show the differences in photon spectra due to the different methods for the photon escape process in Fig. \ref{fig:Method}. Here, we have assumed the standard interstellar medium (ISM) model (constant CSM density) with the macroscopic parameters, the CSM density $n_0$, the initial Lorentz factor $\Gamma_0$, and kinetic energy $E_0$. Throughout this paper, we demonstrate the observed photon flux with a source redshift $z=0.5$, neglecting absorption due to extragalactic background light (EBL). As discussed in \citet{2017ApJ...844...92F}, our time-dependent numerical code produces smoothly curved spectra, in which it is hard to identify the cooling break $\nu_{\rm c}$ and the break frequency $\nu_{\rm m}$ corresponding to the typical emission frequency for electrons with $\gamma=\gamma_{\rm min}$ in the conventional notation.

The red lines in the figure show a significant difference between the spectra obtained from our time-dependent simulation and the simplified analytical method. At $t_{\rm obs}=3000$ s, the photon energy $h \nu_{\rm m}$ corresponding to $\gamma_{\rm min}$ is analytically obtained as $0.14$ eV (here, we adopt $K=4$ for the time correspondence; see equation (\ref{eq:cor}) for the definition of $K$). With the analytical method in \citet{2009ApJ...703..675N} \citep[see also,][]{2022MNRAS.512.2142Y,2024NatAs...8..134A}, the cooling break energy, taking into account the IC cooling is obtained as $h \nu_{\rm c}=13$ eV. Above this energy, $F(\nu) \propto \nu^{-p/2}$ in the analytical formulae, and above $h \hat{\nu}_{\rm c}=8.6$ keV, the Klein--Nishina effect leads to $F(\nu) \propto \nu^{-3(p-1)/4}$. This spectral dimple is smoothed out in the numerical simulation results. First, the photon density in the time-dependent simulation is slightly lower than the value assumed in the analytical formulae, so that the IC flux is lower than the analytical estimate, in which the Compton Y parameter is 7.5 in this case. As a result, the effective cooling frequency becomes higher. In addition, the continuous electron injection in our time-dependent method produces a smoothly curved electron spectrum rather than the sharp break assumed in the analytical method. While those effects lead to the smooth spectra, the overall flux level roughly agrees with the analytical one, in which the normalization factor at $\nu=\nu_{\rm m}$ is taken from \citet{2017ApJ...844...92F}.

As shown in the figure, the slower photon escape in the novel method slightly enhances the effective Compton Y parameter. The differences in the photon flux with the two methods are not larger than a factor of two. 

\section{Models for Shallow Decay}\label{sec:model}

We consider four models in this paper to reproduce the shallow decay behavior in GRB afterglows. The first one is the most promising and well-known model, the energy injection model \citep{2001ApJ...552L..35Z,2006ApJ...642..354Z}. We test two other models in which the microscopic parameters evolve with time \citep{2006MNRAS.369.2059P,2006A&A...458....7I}. The last is the wind model with a relatively low $\Gamma$ \citep{2012ApJ...744...36S,2022NatCo..13.5611D}.
In the first three models, one of the parameters $E_0$, $\epsilon_{\rm e}$, or $\epsilon_B$ evolves with the radius. After the shallow decay phase, the final parameter values are common for those models as $E_0=10^{53}$ erg, $\epsilon_{\rm e}=0.1$, and $\epsilon_B=10^{-3}$. The other parameters are also common as $\Gamma_0=300$, $n_0=1~\mbox{cm}^{-3}$, and $p=2.2$ so that the behaviors in the late stage should be the same.

Note that the parameter evolution in those models required for reproducing the shallow decay lightcurves is not necessarily justified by a well-motivated model, though many studies have been devoted to possible variations of the microscopic parameters with particle-in-cell simulations \citep[e.g.][]{2011ApJ...726...75S,2019ApJ...886...54T,2022ApJ...924L..12P,2022PhRvE.105c5202B,2022ApJ...930L...8V,2024ApJ...963L..44G}. To phenomenologically reproduce the shallow decay behaviors, the parameter evolution is artificially adjusted  until a certain radius $R_{\rm e}$, beyond which the standard afterglow model (constant parameters) is recovered. The radius $R_{\rm e}$ may correspond to the radius within which the CSM plasma is affected by the progenitor activity, which suppresses $\epsilon_{\rm e}$ or $\epsilon_B$. For example, a stronger magnetic field in the upstream may suppress the electron acceleration efficiency \citep{2009ApJ...698.1523S,2010MNRAS.402..321L}, or the heating by the radiation from the progenitor star may smear out the density inhomogeneity, which is seed fluctuation to amplify the magnetic field behind the shock \citep{2011ApJ...734...77I,2019ApJ...886...54T}.

\subsection{Energy Injection Model}
\label{subsec:INJ}

\begin{figure}[t]
    \centering
    \epsscale{1.2}
    \plotone{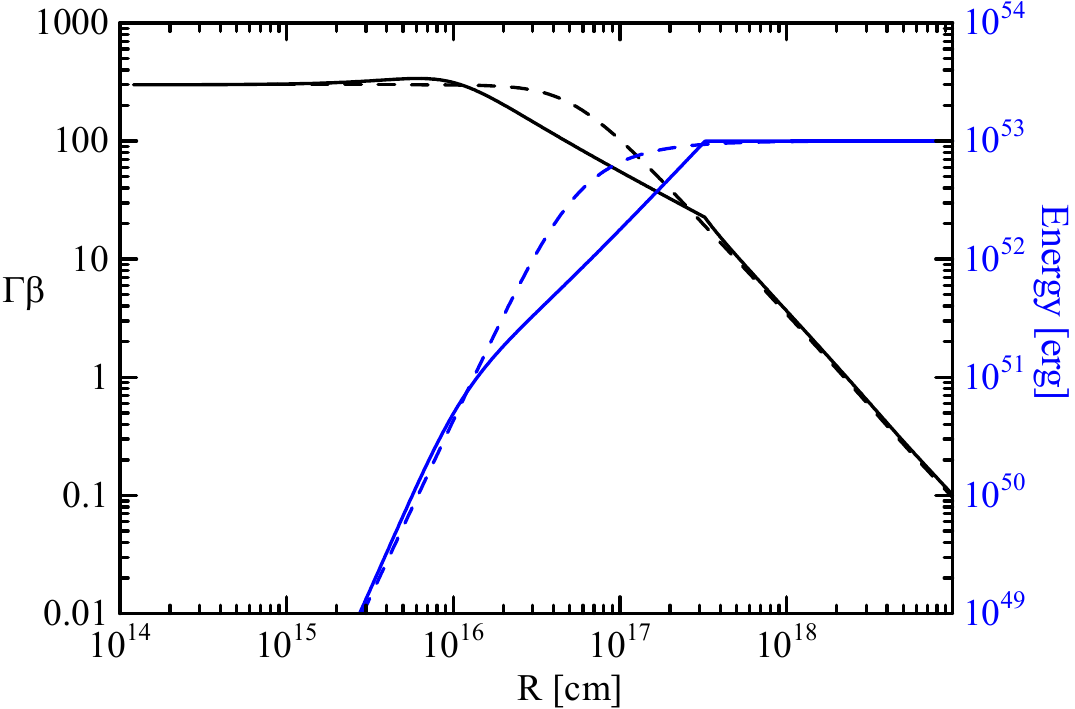}
    \caption{Evolutions of $\Gamma \beta$ (black, left axis) and the total energy injected into the CSM (blue, right axis) in the energy injection model. The dashed lines are behaviors for the standard constant energy model with $E_0=10^{53}$ erg. The parameters are $E_{\rm i}=10^{51}$ erg, $\Gamma_0=300$, $n_0=1~\mbox{cm}^{-3}$, and $R_{\rm e}=3.2\times 10^{17}$ cm.}
    \label{fig:Ene-inj}
\end{figure}

If the central engine acts longer than the prompt emission timescale or the bulk Lorentz factor of ejecta decreases with time, the energy in the forward shock region $E$ increases as the shock front decelerates (the energy injection (EI) model). In most cases, the X-ray band is above $\nu_{\rm c}$ and $\nu_{\rm m}$ in the shallow decay phase. In this case, as discussed in \citet{2001ApJ...552L..35Z}, if we phenomenologically express $E \propto t^s \propto R^s$, the decay index is approximated as $-(2p-2)/3$ or $-(p-2)/2$ for $s=1$ and 2, respectively. To reproduce a shallower decay index than $-0.5$ with $p=2.2$, we adopt $s=2$, which implies a constant luminosity $L_{\rm jet} \propto t^0$ or a $\Gamma$-distribution of $dE/d\Gamma \propto \Gamma^{-5}$ in the long engine-activity model and the decreasing-$\Gamma$ model, respectively.

In this paper, introducing the initial and final energy as $E_{\rm i}$ and $E_0$, respectively, we assume that the total energy increases with $R$ as 
\begin{eqnarray}
E=E_{\rm i}+(E_0-E_{\rm i}) \left(\frac{R}{R_{\rm e}} \right)^2, \quad \mbox{for} \quad R<R_{\rm e}.
\end{eqnarray}
For $R\geq R_{\rm e}$, the total energy is constant $E_0$.

\begin{figure}[t]
    \centering
    \epsscale{2.0}
    \plottwo{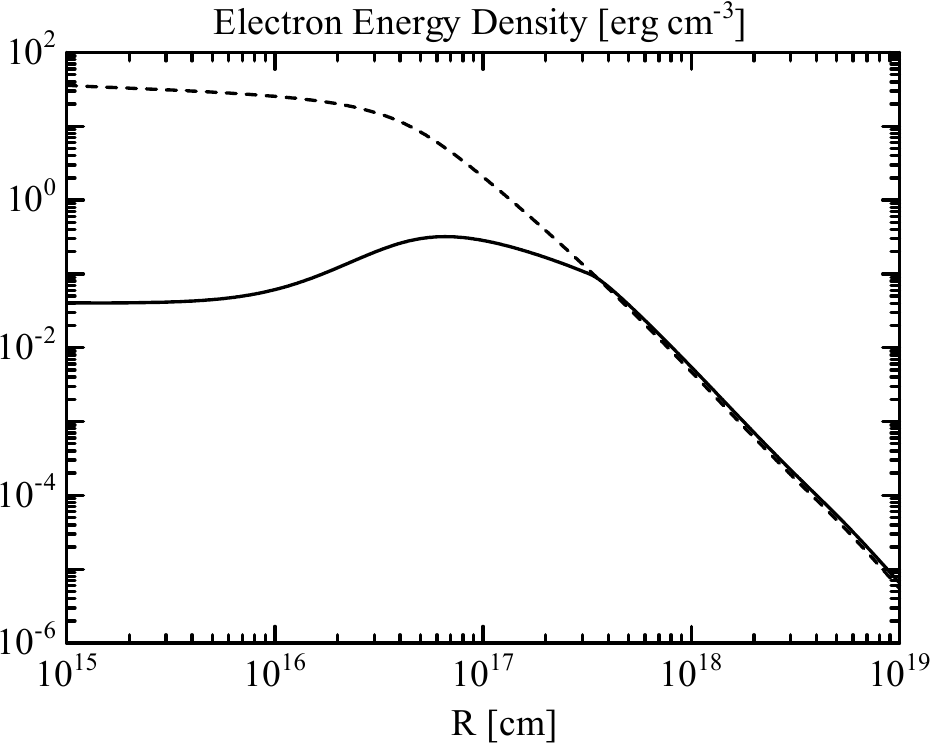}{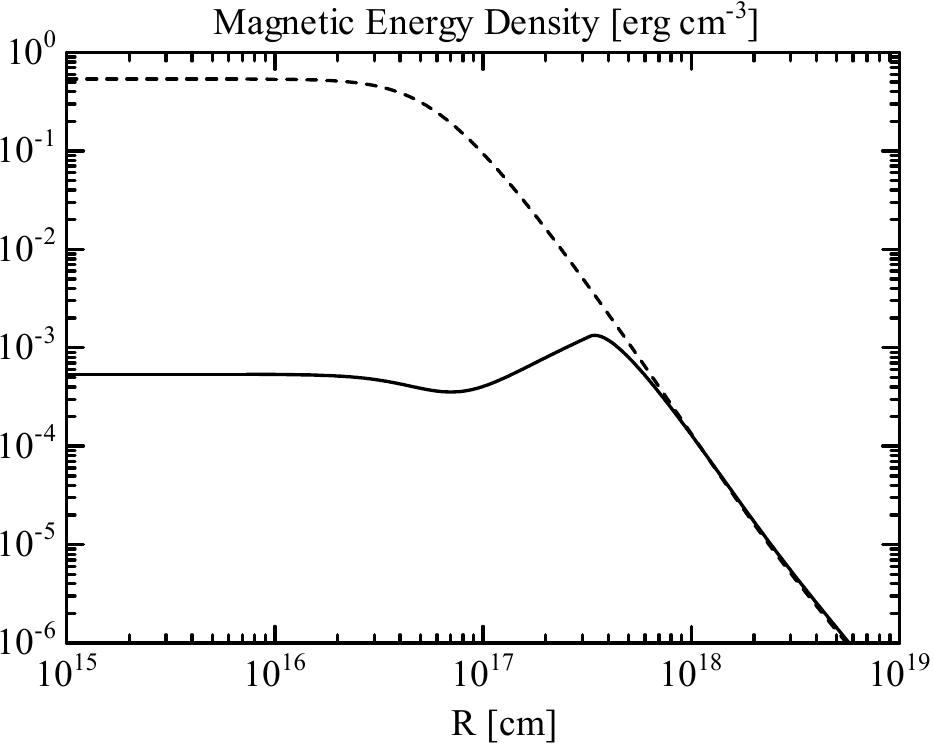}
    \caption{Evolutions of the electron energy density in the evolving $\epsilon_{\rm e}$ model (upper panel) and the magnetic energy density in the evolving $\epsilon_B$ model (lower panel). The dashed lines are behaviors for the standard constant microparameter model ($\epsilon_{\rm e}=10^{-1}$, $\epsilon_B=10^{-3}$). The parameters are $E_0=10^{53}$ erg, $\Gamma_0=300$, $n_0=1~\mbox{cm}^{-3}$, and $R_{\rm e}=3.2\times 10^{17}$ cm.}
    \label{fig:para-ev}
\end{figure}

The evolutions of the bulk 4-velocity $\Gamma \beta$ and the energy injected into the CSM with $E_{\rm i}=10^{51}$ erg and $R_{\rm e}=3.2\times 10^{17}$ cm are shown in Fig. \ref{fig:Ene-inj}. In our simulations, the initial energy is held by the cold ejecta, and the energy is gradually transferred to the CSM via the shock as seen in Fig. \ref{fig:Ene-inj}.
While the simple analytical estimate with $E \propto R^2$ leads to $\Gamma \propto R^{-1/2}$, the numerical results show $\Gamma \propto R^{-0.8}$ for $R<R_{\rm e}$. In Fig. \ref{fig:Ene-inj}, the two break radii for the solid line correspond to the radius where $E=E_{\rm i}$ and $R_{\rm e}$. The solid line roughly connects the two points on the dashed line (constant-$E$ model) at those two radii. Thus, the exact decay index of $\Gamma$ depends on the initial energy and $\Gamma_0$.
For $R>R_{\rm e}$, the bulk Lorentz factor decays following the standard Blandford--Mckee solution $\Gamma \propto R^{-3/2}$.

\subsection{Evolving $\epsilon_{\rm e}$}

Adopting the usual instant control approximation, the flux above $\nu_{\rm c}$ and $\nu_{\rm m}$ decays as $F \propto \epsilon_{\rm e}^{p-1} \epsilon_B^{(p-2)/4} t_{\rm obs}^{(2-3p)/4}$. The one-to-one correspondence approximation with $\Gamma \propto R^{-3/2}$ implies $t_{\rm obs}\propto R^4$. To reproduce the shallow decay with those approximations, we can assume $\epsilon_{\rm e}\propto R^2$, which leads to $F \propto t_{\rm obs}^{-p/4}$. On the other hand, for $\nu_{\rm c}>\nu_{\rm obs}>\nu_{\rm m}$, $F \propto \epsilon_{\rm e}^{p-1} \epsilon_B^{(p+1)/4} t_{\rm obs}^{3(1-p)/4}$. Even in this case, $\epsilon_{\rm e}\propto R^2$ leads to a shallow decay as $F \propto t_{\rm obs}^{(1-p)/4}$. Here, we attempt an evolving $\epsilon_{\rm e}$ (EE) model with a floor value as
\begin{eqnarray}
\epsilon_{\rm e}=\epsilon_{\rm i}+(\epsilon_{\rm f}-\epsilon_{\rm i}) \left(\frac{R}{R_{\rm e}} \right)^2, \quad \mbox{for} \quad R<R_{\rm e}.
\end{eqnarray}
For $R\geq R_{\rm e}$, $\epsilon_{\rm e}$ is constant $\epsilon_{\rm f}$.
We adopt $\epsilon_{\rm f}=0.1$ and $\epsilon_{\rm i}=10^{-4}$.
As shown in Figure \ref{fig:para-ev}, the electron energy density $U_{\rm e}$ is initially suppressed, but the density for $R>R_{\rm e}$ is almost the same as that for the standard constant microparameter model ($\epsilon_{\rm e}=10^{-1}$, $\epsilon_B=10^{-3}$).
Since $U_{\rm e} \propto \epsilon_{\rm e} \Gamma^2$ in the instant control approximation, we can expect $U_{\rm e}\propto R^{-1}$ just before $R_{\rm e}$. Our numerical result shows $U_{\rm e}\propto R^{-0.9}$.

\subsection{Evolving $\epsilon_B$}

In the model with an increasing $\epsilon_B$, the initial magnetic field is relatively low so that $\nu_{\rm c}$ likely becomes significantly higher than the X-ray frequency. For $\nu_{\rm c}>\nu_{\rm obs}>\nu_{\rm m}$, the choice of $\epsilon_B\propto R^4$ is preferable for the shallow decay as $F \propto t_{\rm obs}^{(2-p)/2}$. Similarly to the EE model, we adopt
\begin{eqnarray}
\epsilon_B=\epsilon_{\rm i}+(\epsilon_{\rm f}-\epsilon_{\rm i}) \left(\frac{R}{R_{\rm e}} \right)^4, \quad \mbox{for} \quad R<R_{\rm e}.
\end{eqnarray}
For $R\geq R_{\rm e}$, $\epsilon_B$ is constant $\epsilon_{\rm f}$.
We adopt $\epsilon_{\rm f}=10^{-3}$ and $\epsilon_{\rm i}=10^{-6}$. The evolution of the magnetic energy density $U_B$ is shown in Figure \ref{fig:para-ev}. Similarly to the case in the EE model, the behavior of $U_B$ for $R>R_{\rm e}$ is almost the same as that for the standard constant microscopic parameter model. While the analytical estimate gives $U_B \propto \epsilon_B \Gamma^2 \propto R$ just before $R_{\rm e}$, our numerical result shows $U_B \propto R^{1.1}$.

\begin{figure}[t]
    \centering
    \epsscale{1.2}
    \plotone{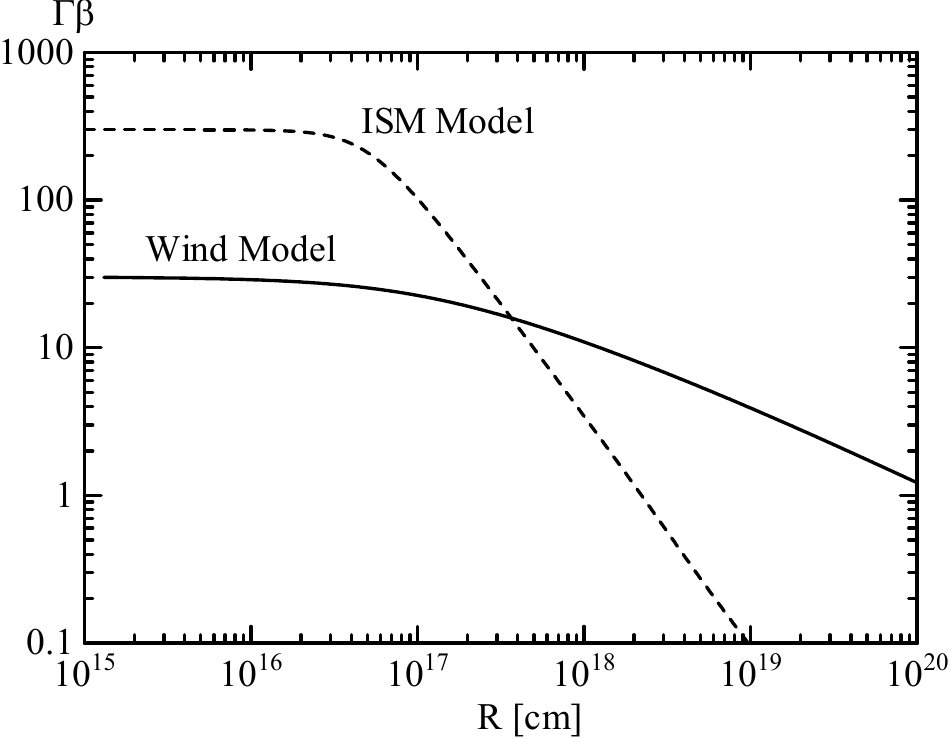}
    \caption{Evolutions of $\Gamma \beta$ for the wind model (solid) and the standard ISM model (dashed; same as the dashed black line in Figure \ref{fig:Ene-inj}). The parameters for the wind model are $E_0=2\times 10^{53}$ erg, $\Gamma_0=30$, and $A_*=0.3$.}
    \label{fig:wind}
\end{figure}

\subsection{Wind Model}

The progenitor stars of GRBs are expected to lose their envelope by stellar winds.
The density profile in the wind model is conventionally parameterized as
\begin{eqnarray}
\rho=5.0 \times 10^{11} A_* \left( \frac{R}{1~\mbox{cm}} \right)^{-2}~\mbox{g}~\mbox{cm}^{-3},
\end{eqnarray}
where $A_*=1$ corresponds to a mass loss rate of $10^{-5} M_\sun~\mbox{yr}^{-1}$ with a wind velocity of $10^3~\mbox{km}~\mbox{s}^{-1}$.
With the mass $M(R)$ within a radius $R$, the deceleration radius $R_{\rm dec}$ is estimated by $E_0=\Gamma_0^2 M(R_{\rm dec})c^2$. In this wind model, we obtain
\begin{eqnarray}
R_{\rm dec}\simeq 2.0 \times 10^{16} A_*^{-1} \left( \frac{E_0}{10^{53}~\mbox{erg}} \right) \left( \frac{\Gamma_0}{30} \right)^{-2} \mbox{cm}.
\end{eqnarray}
The end of the coasting phase for an observer \citep[see e.g.][]{2017ApJ...844...92F} is at
\begin{eqnarray}
t_{\rm dec}&\simeq& (1+z) \frac{R_{\rm dec}}{5c \Gamma_0^2} \\
&\simeq& 150 (1+z) A_*^{-1} \left( \frac{E_0}{10^{53}~\mbox{erg}} \right) \left( \frac{\Gamma_0}{30} \right)^{-4} \mbox{s}.
\end{eqnarray}
During the coasting phase, the analytical estimate leads to $F \propto t_{\rm obs}^{(2-p)/2}$ above the frequencies $\nu_{\rm m}$ and $\nu_{\rm c}$.
Thus, the coasting phase in the wind model can be regarded as a shallow decay phase for an observer.

To elongate the shallow decay phase, we adopt $A_*=0.3$, $E_0=2\times 10^{53}$ erg, and $\Gamma_0=30$, which implies $t_{\rm dec}\simeq 1500$ s for $z=0.5$. This parameter choice is determined to generate a similar flux and lightcurves to those in the other models after the shallow decay phase. As shown in Figure \ref{fig:wind}, the bulk Lorentz factor gradually decreases with radius, and asymptotically behaves as $\Gamma \propto R^{-0.5}$ in the late phase.


\section{Results}
\label{sec:results}

\begin{figure}[t]
    \centering
    \epsscale{1.2}
    \plotone{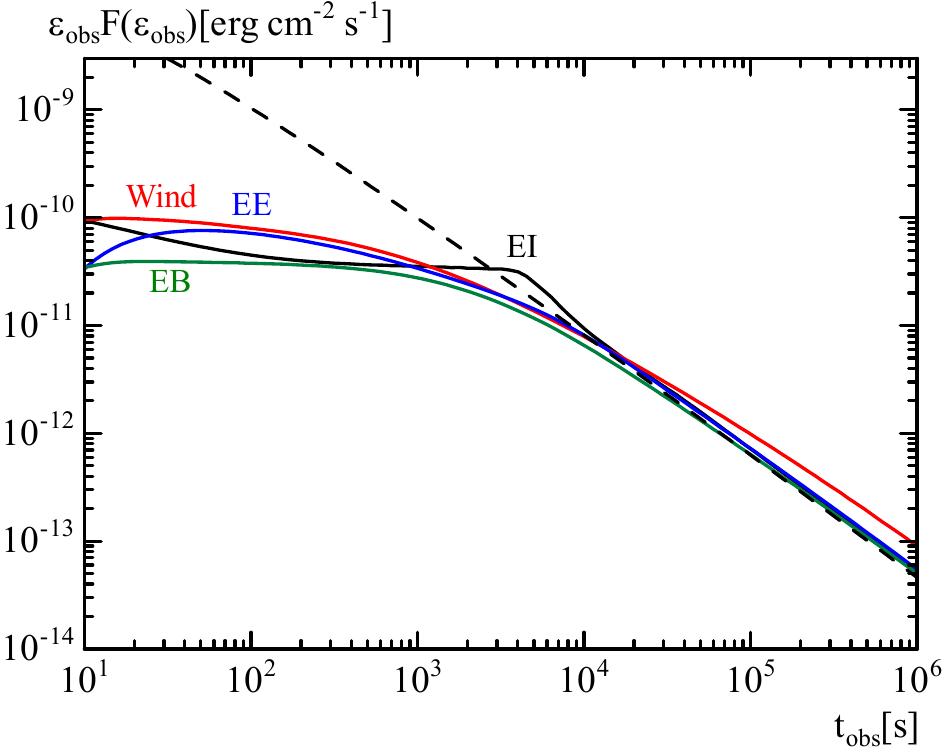}
    \caption{X-ray lightcurves (1 keV) for the energy injection model (EI, solid black), the evolving $\epsilon_{\rm e}$ model (EE, blue), the evolving $\epsilon_B$ model (EB, green), and the wind model (red). The dashed line is for the standard model.}
    \label{fig:XLC}
\end{figure}

Figure \ref{fig:XLC} shows X-ray lightcurves for all the models. Those models well reproduce the shallow decay phase. Considering observational errors, those four lightcurves are almost identical. In the EI model, the Lorentz factor at $R=R_{\rm e}$ is 22.7. The sharp break in the X-ray lightcurve at $t_{\rm obs}=4000$ s implies a factor $K\simeq 8$ for the correspondence
\begin{eqnarray}
t_{\rm obs}&\simeq& (1+z) \frac{R}{Kc \Gamma^2},
\label{eq:cor}
\end{eqnarray}
while the peak time of the afterglow (end of the coasting phase) in the standard model is consistent with $K\simeq 5$ \citep{2017ApJ...844...92F}. Note that the definition of $\Gamma$ is not the shock front Lorentz factor differently from \citet{2021ApJ...923..135D}.

The models with the same $R_{\rm e}$, the EE model, and the evolving $\epsilon_B$ (EB) model, show a more smooth transition from the shallow to the normal decay phase. The lightcurve of the wind model with our parameter choice is similar to that for the EE model. For $R>R_{\rm e}$, the parameter evolutions for the constant CSM models (EI, EE, EB) are the same so that all the lightcurves are almost the same. In the analytical prediction, the decay index for the observation frequency above $\nu_{\rm c}$ and $\nu_{\rm m}$ are common for all the four models. As $E_0$ is higher in the wind model, the flux in the wind model is slightly brighter than the other lightcurves in the normal decay phase.

\begin{figure*}[t]
\begin{minipage}[t]{0.45\linewidth}
    \centering
    \includegraphics[keepaspectratio, scale=0.5]{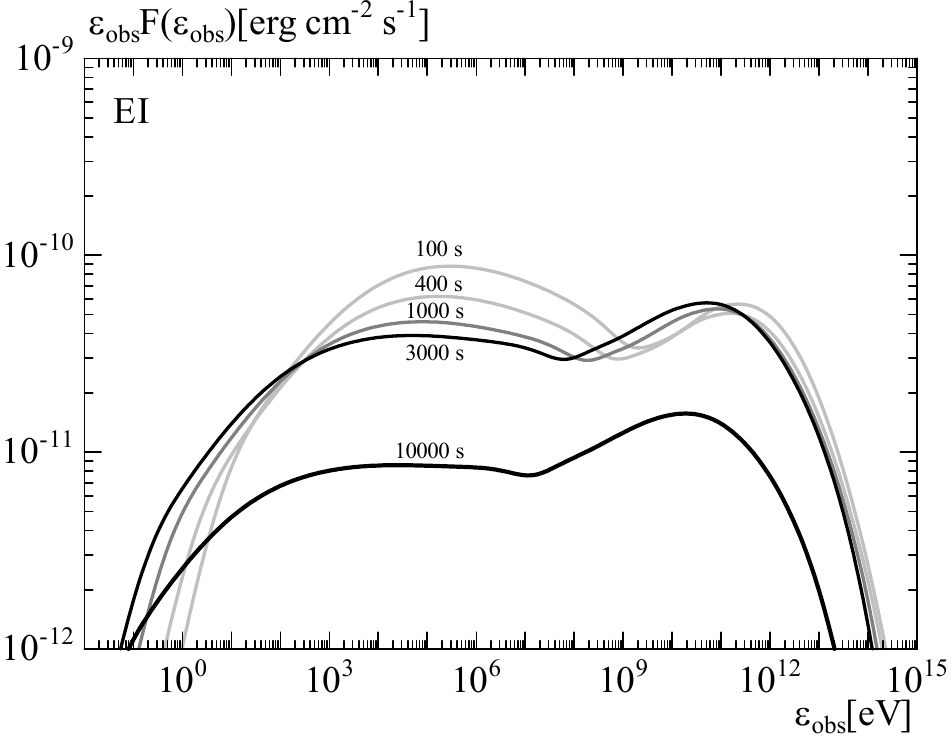}
  \end{minipage} 
  \begin{minipage}[t]{0.45\linewidth}
    \centering
    \includegraphics[keepaspectratio, scale=0.5]{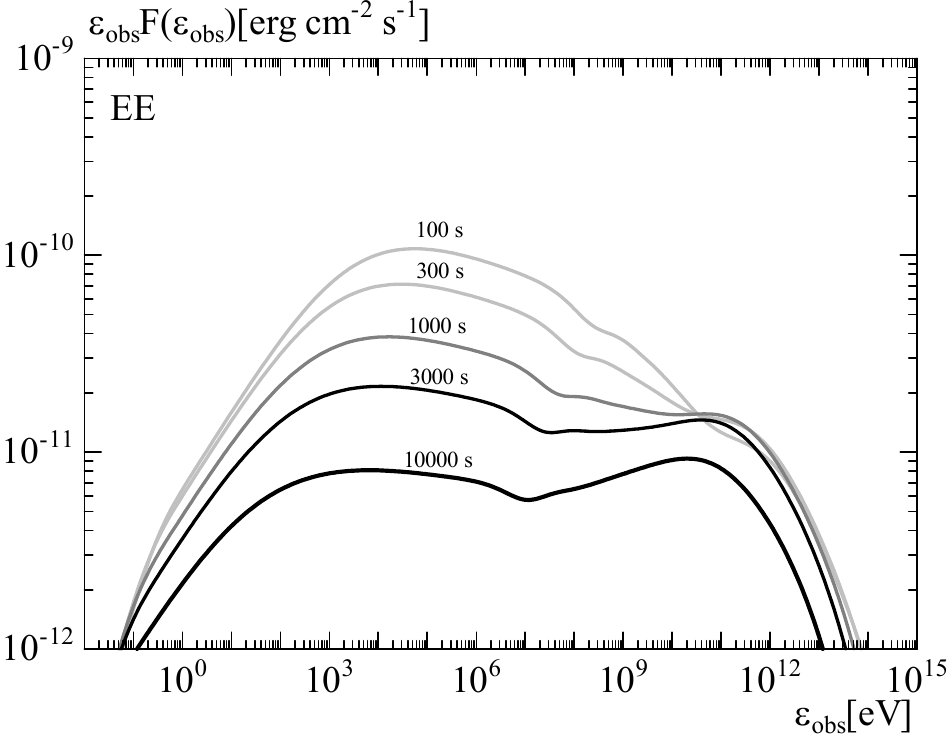}
  \end{minipage} \\
    \begin{minipage}[t]{0.45\linewidth}
    \centering
    \includegraphics[keepaspectratio, scale=0.5]{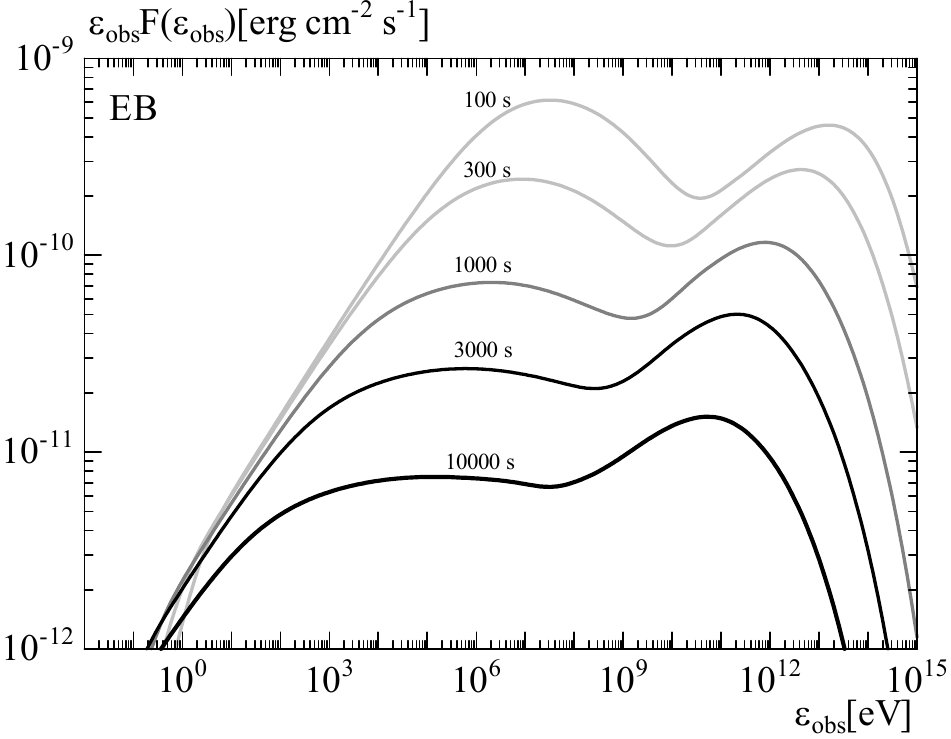}
  \end{minipage} 
    \begin{minipage}[t]{0.45\linewidth}
    \centering
    \includegraphics[keepaspectratio, scale=0.5]{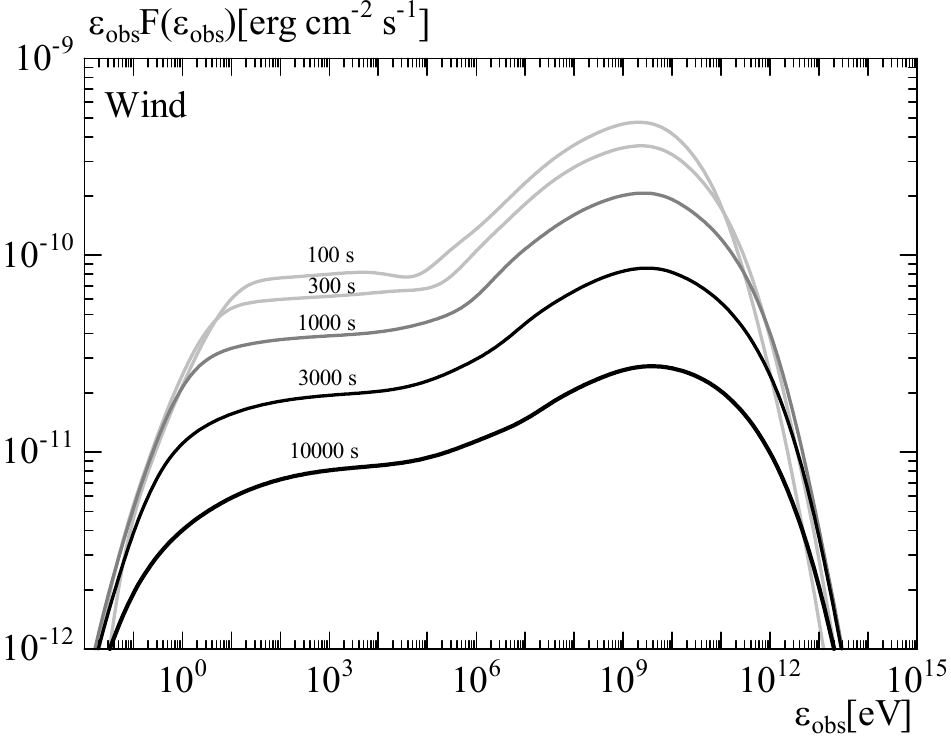}
  \end{minipage}
    \caption{Evolutions of the afterglow spectra from 100 s to 10,000 s for the energy injection model (EI), the evolving $\epsilon_{\rm e}$ model (EE), the evolving $\epsilon_B$ model (EB), and the wind model.}
    \label{fig:spec-evo}
\end{figure*}

Although the X-ray flux in the EI model at the end of the shallow decay phase is slightly brighter than those in the other models, considering observational errors, the four lightcurves are almost identical.

Though the X-ray lightcurves are similar, a characteristic behavior in the evolution of the broadband spectrum is seen for each model (Figure \ref{fig:spec-evo}).
Especially, the differences in the evolution and dominancy of the IC component are distinctly shown in Figure \ref{fig:spec-evo}.
Even with the same $\epsilon_{\rm e}$ and $\epsilon_B$ in the EI and wind models, the gamma-ray flux for the wind model is significantly brighter than the flux for the EI model in the early stage. This is mainly due to the Klein--Nishina effect in the EI model, in which $\gamma_{\rm m} \propto \Gamma$ is higher than that in the wind model. As a result, the spectral peak of the IC component in the wind model ($\sim$ GeV) is lower than those in the other models ($\sim$ TeV). In the EE model, the high initial value of $\epsilon_B/\epsilon_{\rm e}$ suppresses the IC component. In the EB model,
the initial high value of $\epsilon_{\rm e}$ leads to the high flux of the IC component. Differently from the other models, the resultant high $\nu_{\rm c}$ causes a hard X-ray spectrum. Though such hard X-ray spectra in the shallow decay phase are not the majority, several examples have been found \citep{2007ApJ...662.1093W} including
the low-luminosity GRB 171205A \citep{2018A&A...619A..66D,2019Natur.565..324I}.

\begin{figure}
    \centering
    \epsscale{1.2}
    \plotone{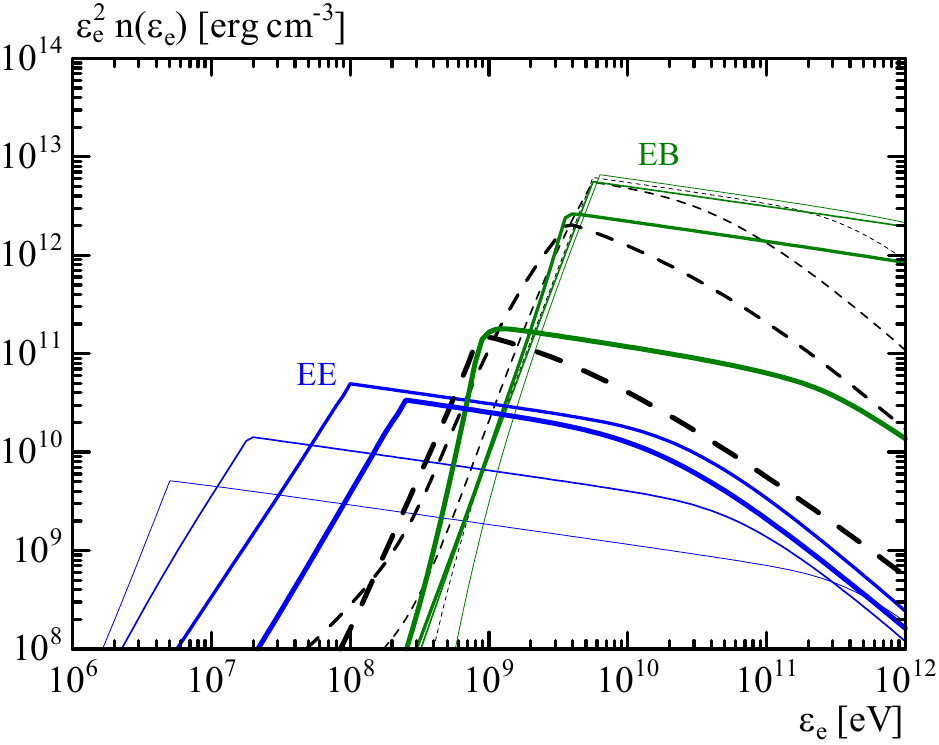}
    \caption{Electron energy evolution for the standard (dashed black), EE (blue), and EB (green) models. The corresponding radii are $1.8\times10^{15}$ cm, $1.8\times10^{16}$ cm, $5.6\times10^{16}$ cm, and $1.8\times10^{17}$ cm from thin to thick lines.}
    \label{fig:ele-ev}
\end{figure}

\begin{figure*}[t!]
\begin{minipage}{0.45\linewidth}
    \centering
    \includegraphics[keepaspectratio, scale=0.5]{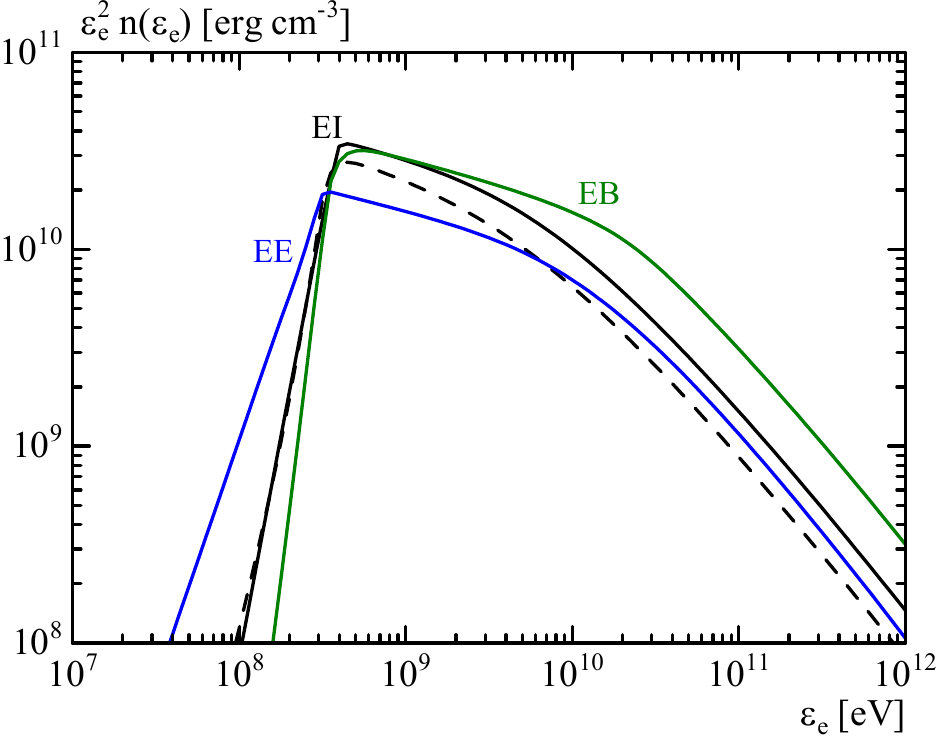}
  \end{minipage} 
  \begin{minipage}{0.45\linewidth}
    \centering
    \includegraphics[keepaspectratio, scale=0.5]{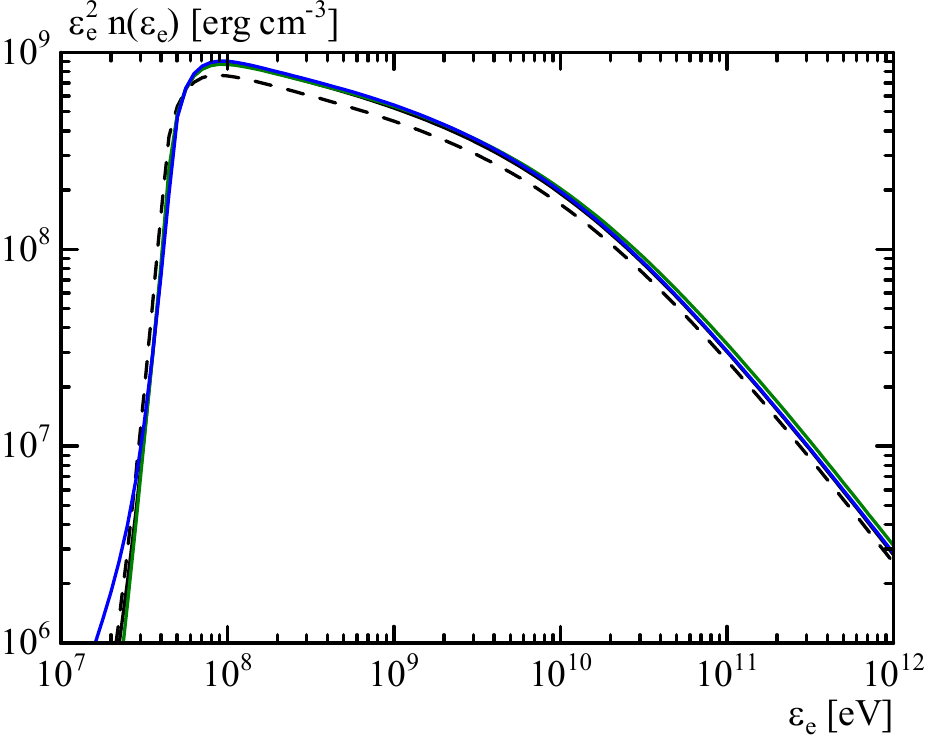}
  \end{minipage} 
    \caption{Electron energy distribution at $R=R_{\rm e}$ ($3.2\times10^{17}$ cm, left) and $R=10^{18}$ cm (right) for the standard (dashed black), EI (solid black), EE (blue), and EB (green) models.}
    \label{fig:ele-spec}
\end{figure*}

In the EI and EB models, the spectral softening is the main reason for the X-ray shallow decay, while the spectral peak significantly decreases with time. Especially in the EB model, the flux in a higher-energy band than X-ray evolves rapidly. On the other hand, the spectral evolutions of the synchrotron component in the EE and wind models are not prominent. The X-ray shallow decay is due to the gradual decay of the synchrotron components maintaining their spectral shapes in those cases.
This gradual evolution in the EE and wind model is reflected in the X-ray lightcurves (Figure \ref{fig:XLC}).

The evolution of the electron energy distributions for the standard, EE, and EB models is shown in Figure \ref{fig:ele-ev}.
In the EE model, the initial low $\epsilon_{\rm e}$ leads to a lower $\gamma_{\rm min}$. With increasing $\epsilon_{\rm e}$, $\gamma_{\rm min}$ also increases, while the cooling break energy decreases with time. In the EB model, the initial low $\epsilon_B$ suppresses the radiative cooling so that the electron spectrum evolves maintaining the hard spectrum.

At $R=R_{\rm e}$, the microscopic parameters and bulk Lorentz factor are common for the EI, EE, and EB models. In this case, the instant control approximation leads to the same electron energy distribution for $R \geq R_{\rm e}$. However, as shown in Figure \ref{fig:ele-spec}, our numerical results show different electron distributions.
In the standard model (dashed line), the radiative energy loss in the early stage appears as the lower electron energy density than that in the EI model, because of the late energy injection at $R \sim R_{\rm e}$. The low radiative efficiency in the EB model leads to a harder spectrum than the other spectra.
The low energy density of the EE model is due to the small $\epsilon_{\rm e}$ in the early stage.
At $R\simeq 3 \times R_{\rm e}$, the electron distributions in the three models are almost identical. At this radius, the information on the initial history of the electron energy distribution almost disappeared.

\begin{figure}
    \centering
    \epsscale{1.2}
    \plotone{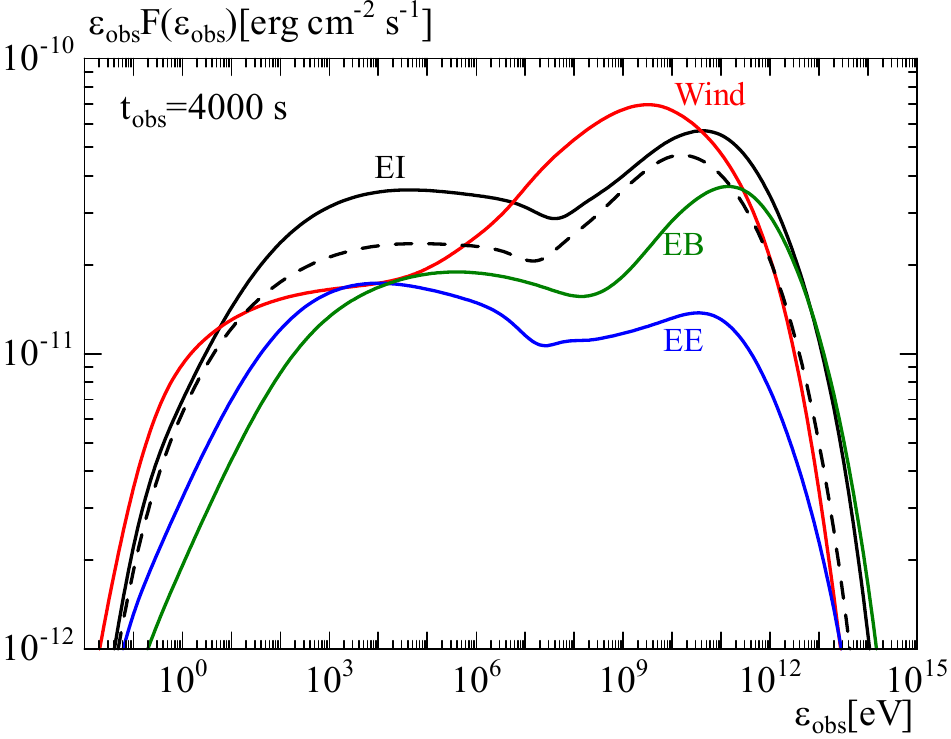}
    \caption{Photon spectra at the end of the shallow decay phase ($t_{\rm obs}=4000$s) for the standard (dashed black), EI (solid black), EE (blue), and EB (green) models.}
    \label{fig:phspec-comp}
\end{figure}

The difference in the electron energy distribution is reflected in the photon spectra (Figure \ref{fig:phspec-comp}). According to the differences in the electron energy distribution at $R = R_{\rm e}$ (Figure \ref{fig:ele-spec}), the fluxes are brighter in the order of the EE, standard, and EI models.
The relatively high ratio of $\epsilon_B/\epsilon_{\rm e}$ in the EE model suppresses the IC component.
As shown in Figure \ref{fig:para-ev}, the magnetic field of the EB model at $R = R_{\rm e}$ is still lower than that in the standard model. In our one-zone treatment, the initial small injection of the magnetic energy in the EB model causes the lower magnetic field at $R = R_{\rm e}$. Though the electron energy density of the EI model at $R=R_{\rm e}$ is close to that of the EB model (Figure \ref{fig:ele-spec}), the photon flux of the EB model is dimmer than that of the EI model. 

\begin{figure*}[t]
\centering
\begin{minipage}[b]{0.65\columnwidth}
    \centering
    \includegraphics[width=1.0\columnwidth]{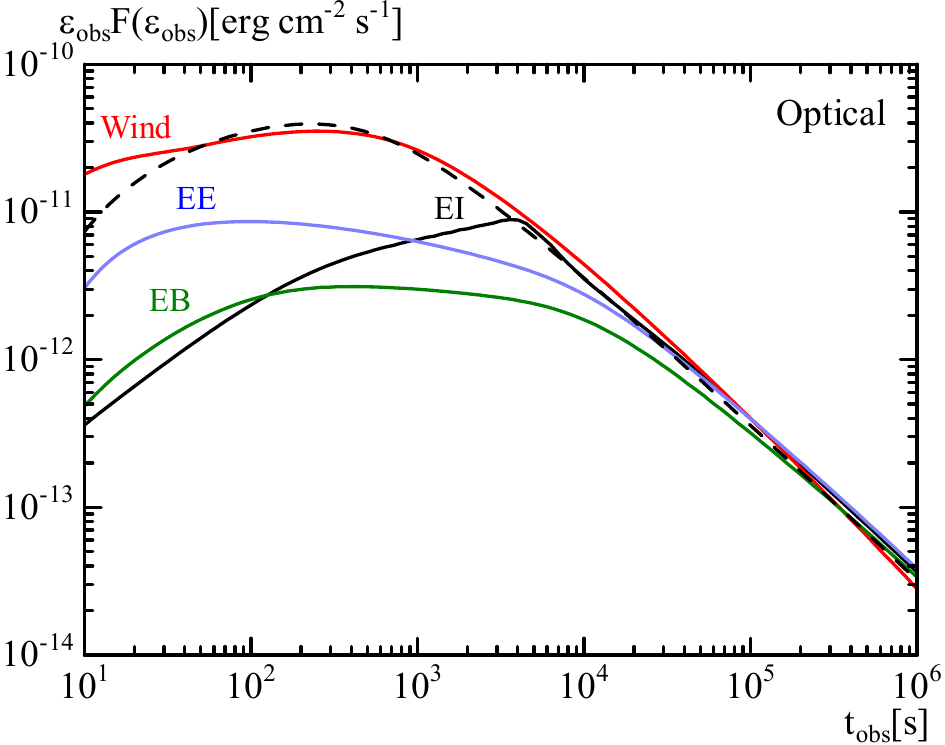}
\end{minipage}
\begin{minipage}[b]{0.65\columnwidth}
    \centering
    \includegraphics[width=1.0\columnwidth]{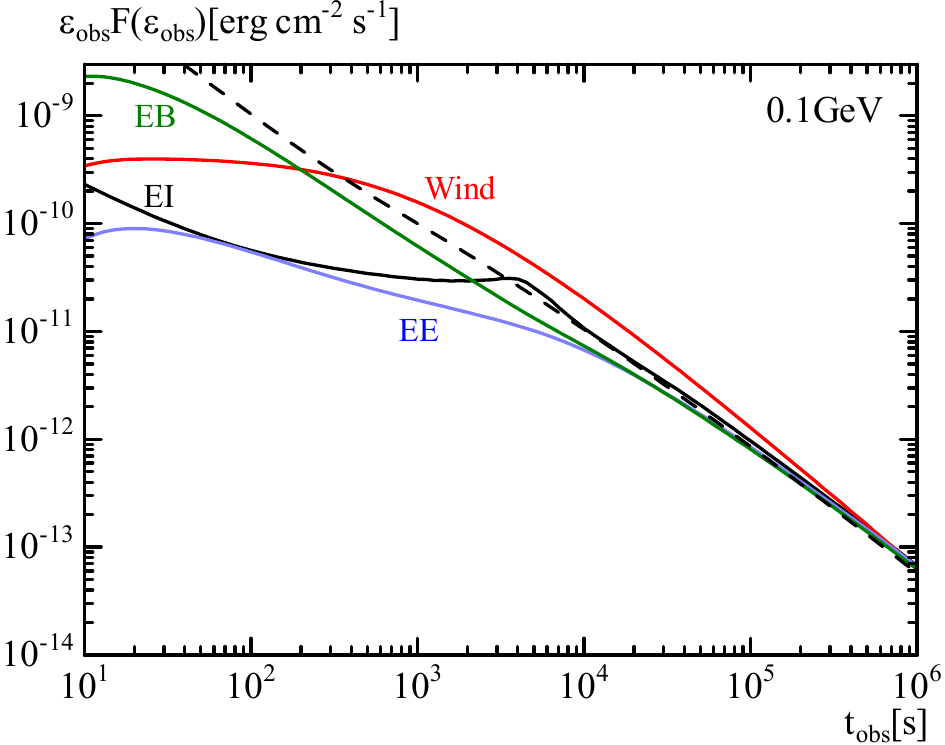}
\end{minipage}
\begin{minipage}[b]{0.65\columnwidth}
    \centering
    \includegraphics[width=1.0\columnwidth]{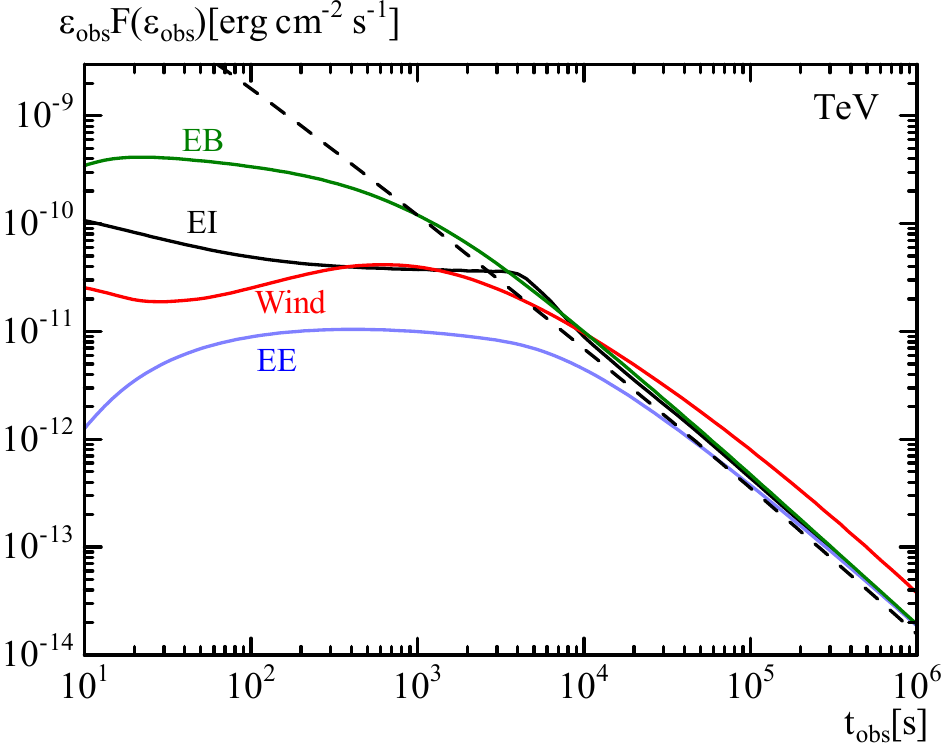}
\end{minipage}
    \caption{Photon lightcurves in optical (left), 0.1 GeV (middle), and TeV (right) bands for the standard (dashed black), EI (black), EE (blue), and EB (green) models.}
    \label{fig:LCs}
\end{figure*}

If we obtain an evolution of the broadband spectrum as shown in Figure \ref{fig:spec-evo}, the discrimination of the models would be easy. However, in most cases, we obtain only lightcurves in limited several frequencies.
In Figure \ref{fig:LCs}, we plot optical and gamma-ray lightcurves for each model.
The optical flux for the EI model characteristically increases until the end of the energy injection. Such lightcurves are very rare, though there are several examples of increasing optical fluxes such as GRB 080710 \citep{2009A&A...508..593K}, whose optical peak time is $\sim 2\times 10^3$ s.

A fraction of shallow decay GRBs show a shallowly decaying optical lightcurve as well \citep[see e.g.][]{2015ApJ...805...13L}. Except for the EI model, the model lightcurves seem consistent with such a shallow decay optical lightcurve, though the fluxes are different.
Even for the standard model, the optical lightcurve shows a similar behavior to the shallow decay one. This is because the optical frequency is below $\nu_{\rm m}$ and $\nu_{\rm c}$. The analytical estimate predicts a growing flux $\propto t^{1/2}_{\rm obs}$ for the slow cooling case in this frequency range \citep{1998ApJ...497L..17S,2017ApJ...844...92F}. The gradual switch from this growing phase to the normal decay phase makes such a shallow-decay-like lightcurve.
The optical break time is not necessarily simultaneous to the X-ray one: some show an earlier break, while some show a later break \citep{2015ApJ...805...13L}. The optical lightcurves for the standard and wind models break earlier, and those for the EE and EB models break later. The variety of the optical break timings can be attributed to the diversity of the shallow decay models.

However, a large fraction of optical lightcurves are consistent with a single power-law decay \citep{2008MNRAS.387..497P,2015ApJ...805...13L}, while the X-ray lightcurve shows a shallow decay behavior. In general, the simultaneous modeling for the X-ray and optical lightcurves with only the forward-shock emission is difficult \citep[e.g.][]{2006MNRAS.369..197F,2009ApJ...702..489R}.
The most promising model for the early optical lightcurves is two-component models such as the forward--reverse shock model \citep[e.g.][]{2014ApJ...781...37P,2014ApJ...785...84J,2019ApJ...879L..26F} and/or spine--sheath model \citep[e.g.][]{2008Natur.455..183R,2023MNRAS.522L..56S}.
In such cases, the contaminated optical lightcurves are not an ideal tool for model discrimination. Therefore, we focus on the gamma-ray afterglow, which may be not polluted by the reverse shock emission \citep[however, see][]{2024NatAs...8..134A}.

As shown in Figure \ref{fig:spec-evo}, except for the wind model, the synchrotron component dominates the 0.1 GeV emission. Figure \ref{fig:LCs} shows that the 0.1 GeV lightcurves should show shallow decay except for the EB model. At $t_{\rm obs}=100$ s, the EB and wind models predict a significantly brighter 0.1 GeV flux than the EI model.

More significant differences are seen in the TeV lightcurves. All the lightcurves show shallow decay, but the flux levels are significantly different during the shallow decay in those models. From only TeV lightcurves, it may be difficult to distinguish the EI and wind models, whose TeV fluxes are similar. However, as shown in Figure \ref{fig:spec-evo}, the wind model predicts a softer spectrum in the TeV energy range. The measurement of the spectral index, which may be disturbed by the EBL absorption though, can distinguish the EI and wind models.

\section{Summary and Discussion}
\label{sec:sum}

We have simulated the emission from the forward shock with the time-dependent code to reproduce the X-ray shallow decay phases. The energy injection (EI) model, evolving $\epsilon_{\rm e}$ (EE) model, evolving $\epsilon_B$ (EB) model, and wind model are considered. Our time-dependent treatment shows that even after the microscopic parameters are common, different parameter evolutions in the early stage affect the spectral shape at the transition phase from shallow decay to normal decay.

The model difference is clearly seen in the evolutions of the broadband spectra. Especially from the evolutions of the IC component, we can distinguish the models. In the EI model, both the synchrotron and IC fluxes do not evolve significantly during the shallow decay phase. If the parameter $\epsilon_{\rm e}$ evolves with radius, the IC component is suppressed in the early stage, and its flux does not largely evolve. In the increasing $\epsilon_B$ model, both the synchrotron and IC components drastically decrease with time, except for the low-energy range (X-ray and optical). In the wind model, the IC component dominates, and its spectral peak is lower than those in the other models.

However, the constraint from GeV--TeV observations has not been significant. While simultaneous optical/IR observations have been implemented, those lightcurves do not usually agree with the model lightcurves. Additional components such as the reverse shock emission are required for optical lightcurves. Though the EB model predicts an unusually hard X-ray spectrum in the early phase, the additional component responsible for optical emission may affect the X-ray spectrum as well. Therefore, we cannot reject or determine a model with the current data.
We expect that future observations with CTA will provide a hint for the mechanism of the enigmatic shallow decay phase.

As there are many parameters in the GRB afterglow physics, we cannot cover all possible cases with different parameter sets in this paper. Furthermore, we do not know typical parameter values in the shallow decay phase because of the lack of GeV--TeV photon detections. To detect TeV photons in the shallow phase of afterglows with CTA, a larger $E_0$ and lower $\epsilon_B$ are ideal. So we expect the simulations in this paper to represent typical examples in future TeV afterglow samples.

\acknowledgments
We appreciate the anonymous referee for the helpful advice. This work is supported by the joint
research program of the Institute for Cosmic Ray Research
(ICRR), the University of Tokyo, and KAKENHI No. 22K03684, 23H04899, and 24H00025.


\bibliography{Bib}{}
\bibliographystyle{aasjournal}

\end{document}